\documentclass[fleqn,10pt]{wlscirep}
\usepackage{epsfig}
\usepackage{color}
\usepackage{graphics}
\usepackage{amsmath}
\usepackage{url}

\title{Organization of fast and slow chromatin revealed by single-nucleosome dynamics}

\author[1]{S. S. Ashwin}
\author[2]{Tadasu Nozaki} 
\author[2,3]{Kazuhiro Maeshima}
\author[1,*]{Masaki Sasai}
\affil[1]{Department of Applied Physics, Nagoya University, Nagoya 464-8601, Japan}
\affil[2]{National Institute of Genetics, Mishima, Shizuoka 411-8540, Japan}
\affil[3]{Department of Genetics, SOKENDAI, Shizuoka 411-8540, Japan}

\affil[*]{masakisasai@nagoya-u.jp}

\begin{abstract}
	Understanding chromatin organization and dynamics is important since they crucially affect DNA functions. 
In this study, we investigate chromatin dynamics by statistically analyzing single-nucleosome movement in living human cells. Bi-modal nature of the mean squared displacement distribution of nucleosomes  allows for a natural categorization of the nucleosomes as fast and slow. Analyses of the nucleosome-nucleosome correlation functions within these categories along with the density of vibrational modes show that the nucleosomes form dynamically correlated fluid regions, i.e., dynamic domains of fast and slow nucleosomes. Perturbed nucleosome dynamics  by global histone acetylation or cohesin inactivation indicate that nucleosome-nucleosome interactions along with tethering of chromatin chains organize nucleosomes into fast and slow dynamic domains. A simple polymer model is introduced, which shows the consistency of this dynamic domain picture. Statistical analyses of single-nucleosome movement provide rich information on how chromatin  is dynamically organized in a fluid manner in living cells.
\end{abstract}
\begin{document}

\flushbottom
\maketitle

\thispagestyle{empty}

\section*{{\it Significan Statement}}
Recent live cell imaging has revealed that chromatin is not a static, rigid structure but is dynamically fluctuating in cells. We study chromatin motion by using single-nucleosome tracking data in living human cells. The distribution of single-nucleosome movement shows a distinct two-peak feature: fast and slow fractions. Nucleosome movement is investigated using statistical analyses and a polymer model to elucidate the mechanism of dynamic organization of fast and slow chromatin.
\\
\\
\\
\section*{Introduction}
Three-dimensional (3D) organization of chromatin in nuclei influences DNA functions such as transcription and replication \cite{Bickmore2013, Cardoso2012, Hubner2013, Maeshima2016}, and hence has been a focus of  intensive investigation. In particular, the high-throughput chromosome conformation capture (Hi-C) approaches have revealed frequent chromatin-chromatin interactions within topologically associating domains (TADs) of several $10^2$~kb in size   \cite{Dekker2015} or loop domains formed by looped chromatin chains of  $\sim 10^2$~kb \cite{Rao2014}; these domains have come to be regarded as basic structural units of the genome chromatin. However, the single-cell Hi-C measurements \cite{Nagano2017,Tan2018} and the microscopic observations with fluorescence in situ hybridization (FISH) \cite{Bintu2018, Finn2019} have shown that positions of domain boundaries along the sequence are primarily different from cell to cell, showing large structural fluctuation in domain formation. Therefore, a fundamental question arises on how chromatin domains are formed with intense fluctuation and how such fluctuating domain formation affects gene regulation.

In order to answer this question on fluctuating chromatin organization, it is important to analyze chromatin movement in living cells. 
 Highly dynamic chromatin movement has been so far observed with live-cell imaging studies by using the LacO/LacI-GFP and related methods \cite{Marshall1997, Heun2001, Chubb2002, Levi2005, Hajjoul2013,Germier2017}, and by more recent  CRISPR/dCas9-based methods \cite{Chen2013, Gu2018, Ma2019} and the single-nucleosome tracking techniques \cite{Hihara2012, Nozaki2013, Nozaki2017, Nagashima2019}. In particular, Nozaki et al. observed single-nucleosome movement in living human cells in a genome-wide manner and showed that nucleosomes are clustered in the nuclei to form chromatin domains \cite{Nozaki2017}. Here, we write displacement of the $i$th nucleosome during time period $t$ as $\delta {\rm {\bf r}}_i(t)={\rm {\bf r}}_i(t+t_0)-{\rm {\bf r}}_i(t_0)$. When the nucleosome $i$ belongs to the domain $\alpha$, we can write $\delta {\rm {\bf r}}_i(t)=\delta {\rm {\bf r}}_i^{\rm intra}+\delta {\rm {\bf r}}_{\alpha}$ for $t$ shorter than the timescale of domain formation/dissolution, where $\delta {\rm {\bf r}}_i^{\rm intra}$ represents the intra-domain displacement of the nucleosome and $\delta {\rm {\bf r}}_{\alpha}$ is the center of mass movement of the domain. Nozaki et al. exemplified cases that single-nucleosome movement is correlated with the domain movement for $t\sim 0.1$--1\,s \cite{Nozaki2017}, suggesting domain movement is dominant with $|\delta {\rm {\bf r}}_{\alpha}|$ being sufficiently large. 
 Theoretical polymer models also highlighted the effects of domain movement on chromatin dynamics \cite{Pierro2018, Shi2018, Shinkai2016}; in the model of Pierro et al. \cite{Pierro2018}, motions of different chromatin loci within the same domain are correlated with each other, showing significance of $|\delta {\rm {\bf r}}_{\alpha}|$. Thus, quantitative analysis of chromatin dynamics is a key to understand how chromatin domains are dynamically organized.

An important aspect of chromatin dynamics is their heterogeneity. Here,  $M_i(t)=\left< \delta {\rm {\bf r}}_i(t)^2\right>_{t_0}$ is mean square displacement (MSD) of the chromatin locus $i$ and $\left< \cdots\right>_{t_0}$represents average over $t_0$. In the previous study, $M_i$ at particular loci of chromosomes \cite{Hajjoul2013, Gu2018, Bronstein2009}, and $\bar{M}$ averaged over a wide region of the nucleus \cite{Zidovska2013} or over the genome-wide single-nucleosome ensemble \cite{Hihara2012, Nozaki2013, Nozaki2017, Nagashima2019} were examined, showing chromatin movement is sub-diffusive as $M \sim t^{\beta}$ with $\beta<1$, where various different values of  $0.3 \lesssim \beta < 1$ were reported, suggesting diversity of chromatin movement. In the polymer model of Shi et al. \cite{Shi2018}, the $\beta$ value depends on whether the domain of the calculated locus is near the surface or in the interior of a chromosome structure. Shinkai et al. \cite{Shinkai2016} argued that $|\delta {\rm {\bf r}}_i^{\rm intra}|$ is smaller in compact heterochromatin-like domains, resulting in the smaller $\beta$. Thus, the observed and calculated $\beta$ suggested that chromatin motion depends on interactions and environments of individual domains; therefore, chromatin movement is heterogeneous in a complex genomic structure. Indeed, heterogeneous distribution of movement was observed in fluorescent images of living mammalian cells using single-nucleosome tracking as ``chromatin heat map'' \cite{Nozaki2017} and by flow-field monitoring \cite{Shaban2019}. Thus, it is important to analyze heterogeneity in chromatin dynamics, which gives a clue to elucidate interactions and varied local environments of the chromatin domains.

In the present study, we statistically analyze heterogeneity in chromatin dynamics by using the live-cell imaging data of Nozaki et al. \cite{Nozaki2017}. These data are single-nucleosome trajectories obtained by tracking fluorescent images of nucleosomes in a thin layer ($\sim200$--250 nm thickness) of the HeLa cell nucleus. An example image is shown in \textcolor{blue}{Fig.\,S1} and \textcolor{blue}{Movie S1}. See the \textcolor{blue}{SI text} for more details. Using single-nucleosome trajectories, we extract distributions of MSD of individual nucleosomes, which allows for the nucleosome characterization as fast and slow. Based on this classification, features of nucleosome packing in chromatin domains are inferred by analyzing auto- and pair-correlations of nucleosome movement and by comparing cells in different conditions. Chromatin regions in which single nucleosomes show correlated movement are referred to as fast and slow dynamic domains. A minimal polymer model is introduced to elucidate mechanics governing these domain organization.

\section*{Results}
\subsection*{Fast and slow fractions of nucleosomes}

\begin{figure}
\centering
\includegraphics[width=0.6\linewidth]{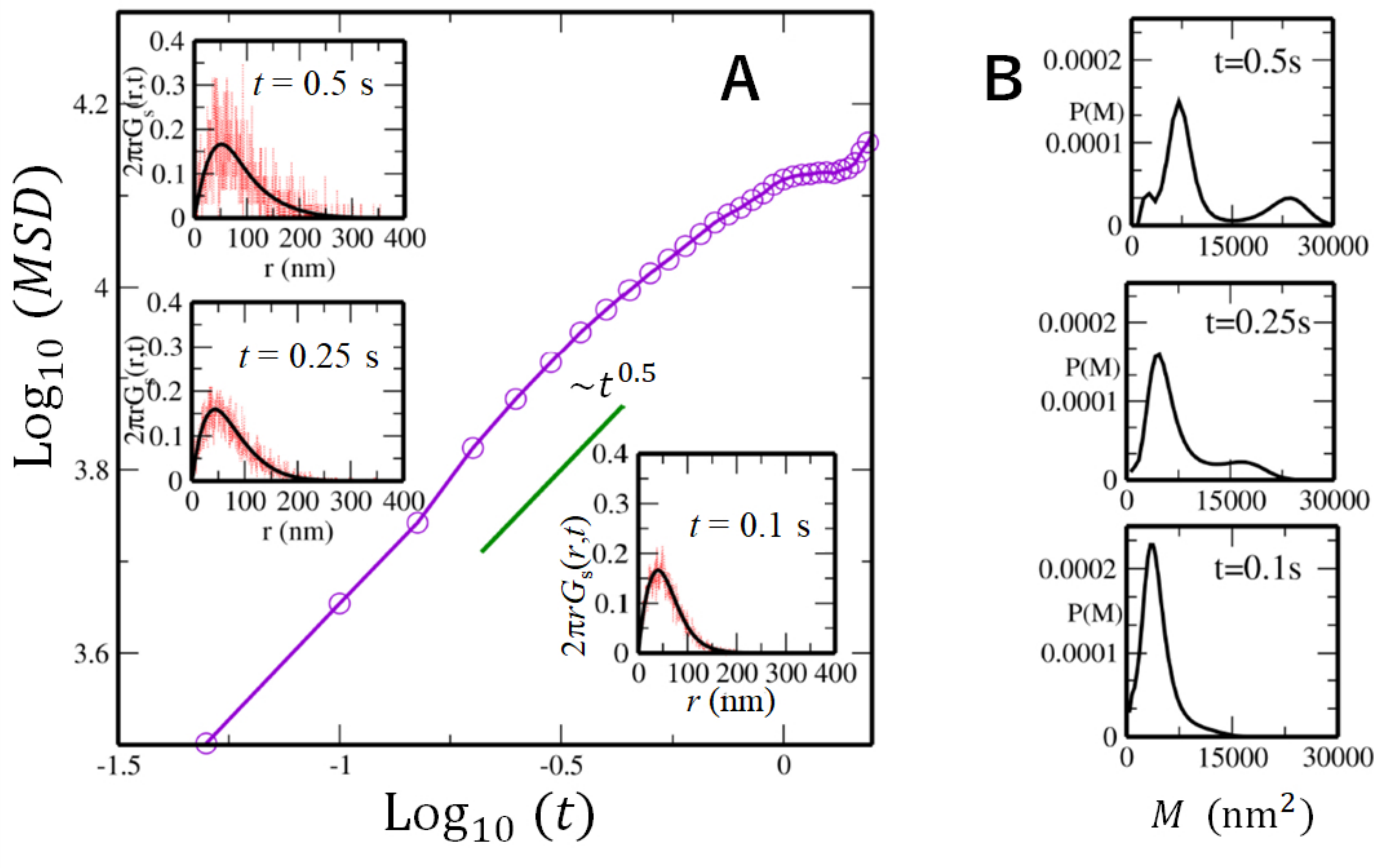}
\caption{
Mean square displacement (MSD) of nucleosome movement observed in live-cell imaging of an example cell. (A) The MSD $\bar{M}$ averaged over nucleosomes is plotted as a function of time. In insets, the self-part of the van Hove correlation function $2\pi rG_s(r,t)$ reproduced from $P(M,t)$ using Eq.~1 (black) is superposed on the one obtained from the observed trajectories of single nucleosomes (red) at $t=0.1$\,s, $t=0.25$\,s and $t=0.5$\,s. (B) The distribution $P(M,t)$ of the MSD of single nucleosomes at $t=0.1$\,s, $t=0.25$\,s and $t=0.5$\,s.
}
\end{figure}

Shown in Fig.~1A is the average MSD of nucleosomes,  $\bar{M}=\left< M_i \right>$, where $\left< \cdots \right>$ is the average taken over $i$ and along the observed trajectories \cite{Nozaki2017}. 
 From Fig.~1A, we find that the average movement of nucleosomes is sub-diffusive for $t<1$\,s with $\bar{M}\sim t^{0.5}$ though individual nucleosomes move with different exponents from 0.5 as shown below. For $t\sim 1$\,s, $\bar{M}$ tends to saturate, suggesting nucleosomes are caged in finite  regions, and for $t>1$\,s, nucleosomes begin to diffuse with $\bar{M}\sim t$. Here, we focus on the timescale $t<1$\,s, where the data is experimentally well sampled by Nozaki et al. \cite{Nozaki2017}. See the \textcolor{blue}{SI text} and \textcolor{blue}{Fig.~S2} for the details of sampling the data.

A remarkable feature is the diverse distribution of MSD of individual nucleosomes, which is captured by the distribution function, $P(M,t)=\left< \delta(M-M_i(t)) \right>$. Due to the short lifetime of observed fluorescence of  single nucleosomes, the  individual nucleosome MSD data is insufficient to provide for a clear $P(M,t)$. However, this problem is overcome by using the iterative algorithm of Richardson and Lucy (RL) \cite{Richardson1972, Lucy1974} to derive the smooth distribution from the noisy data. From the observed data, we first calculate the self-part of the van Hove correlation function (vHC), $G_s(r,t)=A_s\left< \delta(r-|{\rm{\bf r}}_i(t+t_0)-{\rm{\bf r}}_i(t_0)|)\right>$, where ${\rm{\bf r}}_i$ is the projected coordinate of the $i$th nucleosome on the 2D imaging plane and $A_s$ is a constant to normalize $G_s$ as $\int d^2{\rm {\bf r}}G_s(r,t)=1$. The calculated vHC is shown at $t=0.1$\,s, 0.25\,s and 0.5\,s in insets of Fig.~1A. $G_s$ is expanded in Gaussian bases, $q(r,M)=(1/{\pi M)\exp( -r^2/M})$, as
\begin{equation}
G_s(r,t)=\int dM P(M,t)q(r,M). 
\end{equation}
Given a noisy estimate of $G_s (r,t)$, $P(M,t)$ is extracted as coefficients of expansion using the RL iterative scheme. See the {\it Methods} section. The RL algorithm has been extensively used in image processing \cite{Dey2006, Laasmaa2011} and also in monitoring diffusion of liposomes in a nematic solution \cite{Wang2012} and particles in simulated supercooled liquids \cite{Sengupta2014, Bhowmik2016, Bhowmik2018}. In \textcolor{blue}{Figs.~S3--S5}, we show that this iterative method works well for evaluating the MSD distribution  in example polymer systems.

\begin{figure}
\centering
\includegraphics[width=0.6\linewidth]{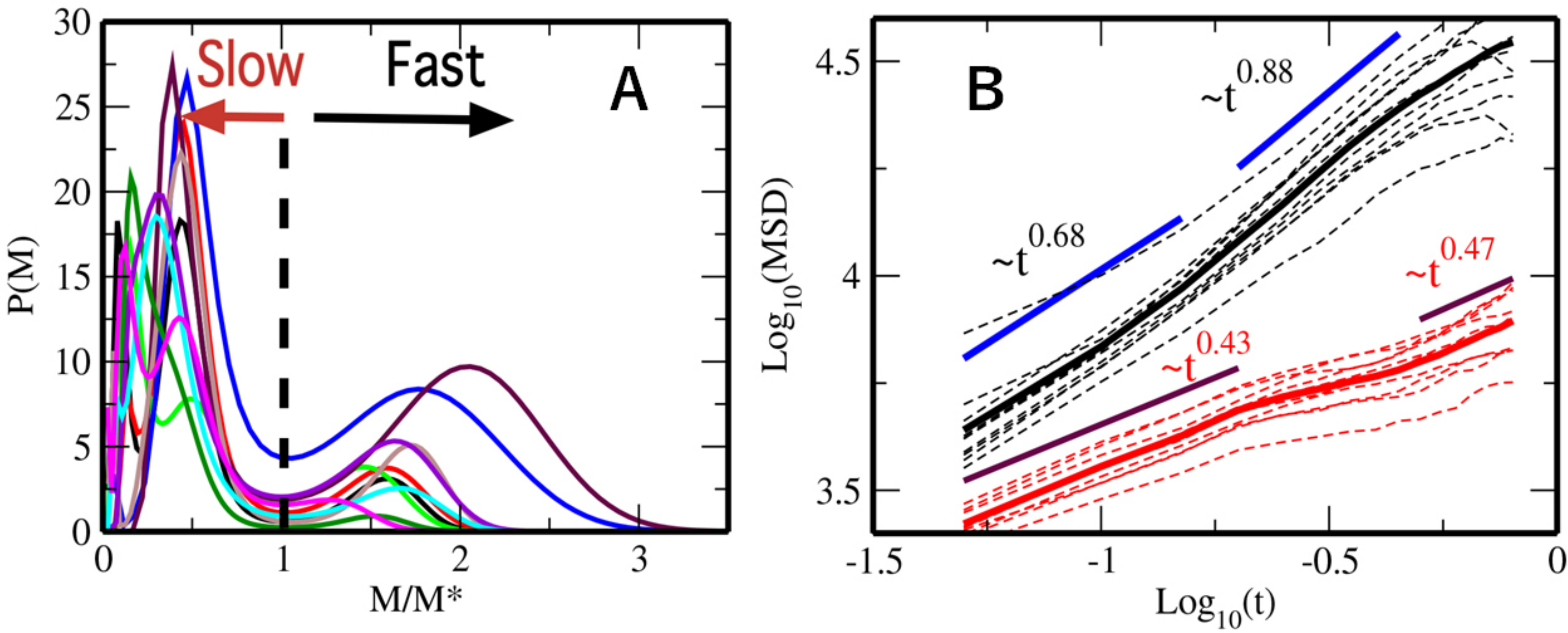}
\caption{
Fast and slow nucleosomes. (A) The distribution of MSD of single nucleosomes, $P(M)=P(M,0.5~{\rm s})$, is plotted for 10 cell samples as functions of $M/M^*$, where $M^*$ is $M$ at the minimum between two peaks of $P(M)$. (B) The MSD averaged over fast nucleosomes, $\bar{M}_f$ (black), and the MSD averaged over slow nucleosomes, $\bar{M}_s$ (red), are shown for 10 individual cells (dashed lines) and the average over 10 cells (real lines).
}
\end{figure}

The MSD distribution, $P(M,t)$, obtained from the RL scheme is shown in Fig.~1B. At $t=0.1$~s, the distribution shows a single peak, but it splits into two peaks with increasing time. This bimodal feature is distinct at $t=0.5$\,s, which allows for a characterization of nucleosomes as fast and slow. 
Mobility of nucleosomes varies from cell to cell, but the functional form of $P(M,t)$ becomes similar to each other when $M$ is scaled in a suitable way. In Fig.~2A, we plot $P(M,t=0.5~{\rm s})$ for the 10 cells we examined as functions of the scaled MSD, $M/M^{*}$, where $M^{*}$ is defined as $M$ at the minimum of $P(M,0.5~{\rm s})$.
Thus, we define fast (slow) nucleosomes as ones showing $M_i(0.5\,{\rm s})\ge M^*$ ($M_i(0.5\,{\rm s})< M^*$). 
Then, the full vHC can be written in terms of sum of the vHC of the fast and slow nucleosomes as $G_s (r,t)=G_s^f (r,t)+G_s^s (r,t)$ with $G_s^f(r,t)=\int_{M^*}^{\infty} dMP(M,t)q(r,M)$ and $G_s^s(r,t)=\int_0^{M^*}dM P(M,t)q(r,M)$. 
See \textcolor{blue}{Fig.~S4} for the validity of this decomposition. 
With this characterization, we separately calculate the average MSD by   $\bar{M}_a=\int r^2 G_s^a(r,t)d^2{\bf r}$ for the fast ($a=f$) and slow ($a=s$) nucleosomes as shown in Fig.~2B. When we fit the MSD as $\sim t^\beta$, the exponent is $\beta =0.69$--0.88 for the fast nucleosomes and $\beta =0.44$--0.47 for the slow nucleosomes. This suggests that fast and slow nucleosomes move in different physical mechanisms.

\subsection*{Fast and slow dynamic domains}

In order to understand the organization and underlying mechanisms that govern the dynamics of fast and slow movements, we analyze their temporal and spatial correlations. Shown in Fig.~3A are auto-correlations of displacement of nucleosomes;
\begin{equation}
\eta^a(t)= \left[ \left< {\rm {\bf v}}_i(t+t_0)\cdot{\rm {\bf v}}_i(t_0)\right>_{i\in a}/ \left< {\rm {\bf v}}_i(t_0)^2\right>_{i\in a}\right]_{\rm cell},
\end{equation}
where  $\left< \cdots \right>_{i\in a}$ with $a=f$ ($a=s$) is the average over $t_0$ and over the fast (slow) nucleosomes and  $[\cdots ]_{\rm cell}$ is the average over 10 cells. ${\bf v}_i (t)$ is the displacement vector of the nucleosome $i$, ${\rm {\bf v}}_i(t)=\left({\rm {\bf r}}_i(t+\delta t)-{\rm {\bf r}}_i(t)\right)/\delta t$, with  $\delta t =0.05$\,s. Fig.~3A shows that $\eta^a(t)$ changes its sign for the first time at $t=t^*\approx \delta t$ for both $a=f$ and $s$, which indicates back scattering from the neighboring nucleosomes. The similar viscoelastic behavior has been seen in a polymer model of chromosomes \cite{Pierro2018}. For $t>t^*$, $\eta^a(t)$ shows a damped oscillation with a period $\sim 2t^*$. From $t^*$, a typical distance, $d_a$, for nucleosomes to traverse between successive collisions is estimated as $d_a=\int rG_s^a(r,t^*)d^2{\bf r}$, showing $d_f\approx 74$--88 nm for the fast nucleosomes and $d_s\approx 36$--43 nm for the slow nucleosomes.

\begin{figure}
\centering
\includegraphics[width=0.6\linewidth]{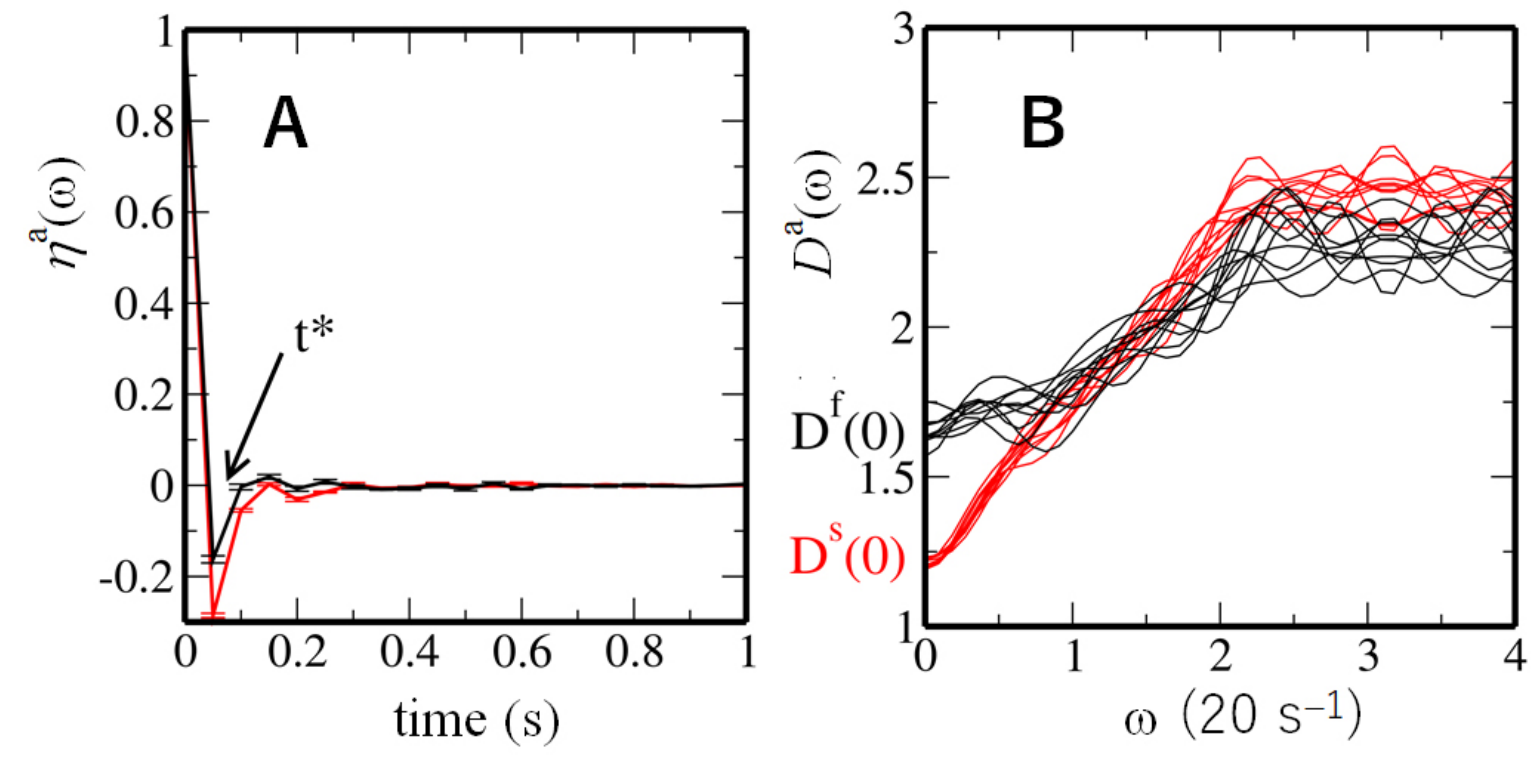}
\caption{
Auto-correlation functions of displacement of single nucleosomes and the density of vibrational modes. (A) The auto-correlation function of single nucleosome displacement, $\eta^a(t)$, is plotted as a function of $t$. Bars show the standard errors among 10 cells. (B) The density of vibrational modes, $D^a(\omega)$, is plotted as a function of frequency $\omega$ for 10 cells. 
In A and B, curves are plotted for fast ($a=f$, black) and slow ($a=s$, red) nucleosomes.
}
\end{figure}

We can approximately regard ${\bf v}_i (t)$ as a velocity vector; then, the Fourier transform of $\left< {\rm {\bf v}}_i(t+t_0)\cdot{\rm {\bf v}}_i(t_0)\right>_{i\in a}$, denoted here by $D^a(\omega)$, is the approximate density of vibrational modes (Fig.~3B). It is interesting to note that $D^a (0)$ is finite, which is a signature of liquid-like behavior: $D^a(0)\neq 0$ represents self-diffusion in a fluid state while $D^a(0)=0$ in an
amorphous solid state \cite{Lin2003}. $D^f (0)>D^s (0)$ shows that the fast nucleosomes are more fluid, and $D^f(\omega) < D^s(\omega)$ for the large $\omega$ shows that movement of the slow nucleosomes is more constrained. Nature of chromatin packing at the 30 nm scale has been under debate whether the regular 30-nm fibers \cite{Finch1976} exist or not,  with emerging evidence for the fluid movement of chromatin in this length scale \cite{Maeshima2016, Maeshima2019}. Here, the auto-correlations of displacement showed that the nucleosomes are back scattered at the 30 nm length scale along with the finite vibrational state density of  $D^a(0)\neq 0$, further providing an evidence for the fluid nature of chromatin at this length scale.

\begin{figure}
\centering
\includegraphics[width=0.6\linewidth]{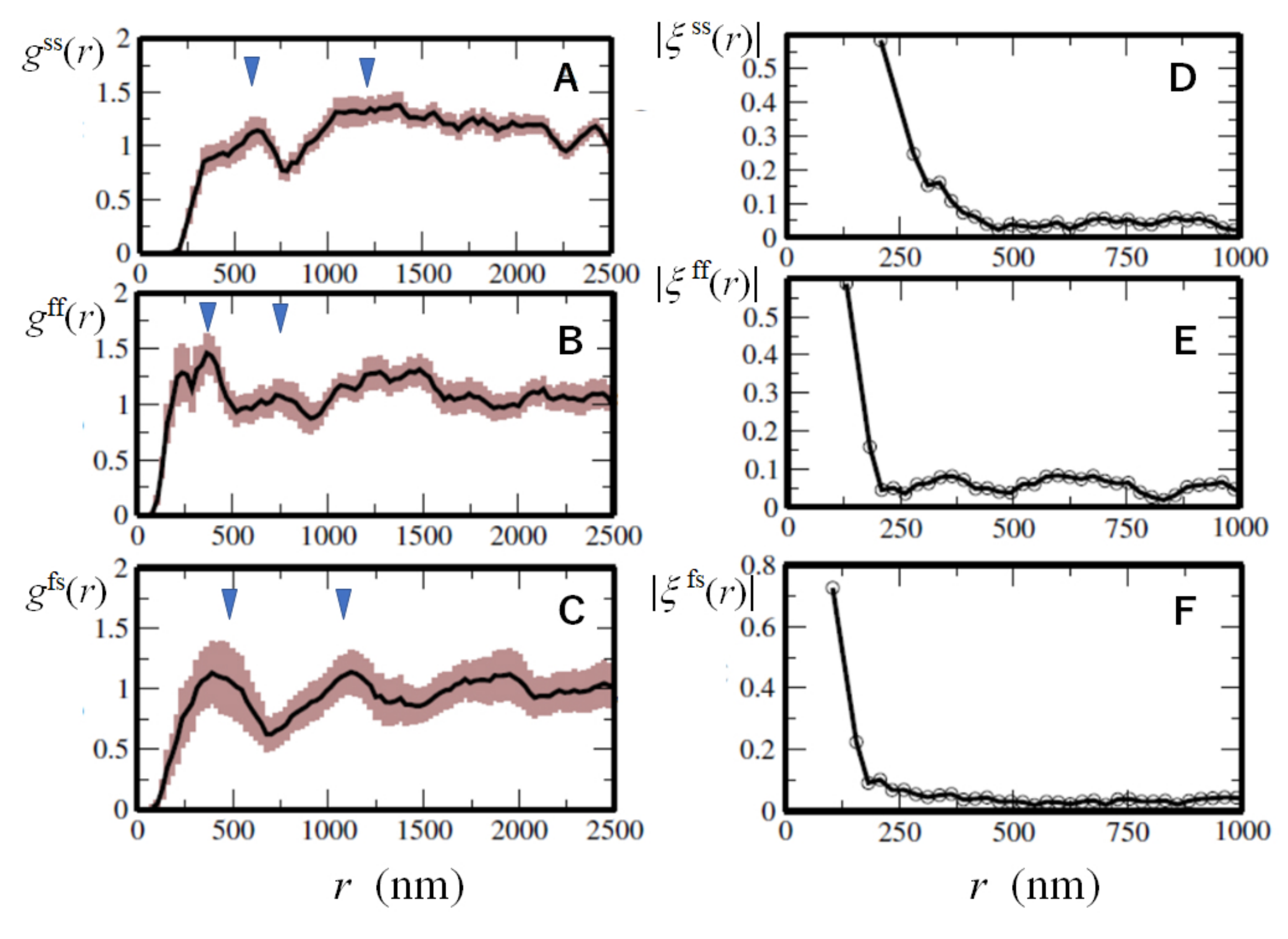}
\caption{
Pair-correlation functions of position and displacement of single nucleosomes. (A--C) Pair-correlations of position, i.e., the radial distribution functions, $g^{ab}(r)$, of single nucleosomes. Triangles show the distances, $D^{ss}$ and $2D^{ss}$ (A), $D^{ff}$ and $2D^{ff}$ (B), and $D^{fs}$ and $D^{fs}+D^{ss}$ (C). 
Width of brown shaded area shows the standard errors among 10 cells. (D--F) Profile functions, $|\xi^{ab}(r)|$, of pair-correlation of displacement of single nucleosomes. Curves are shown with $ab=ss$ for the slow-slow correlation (A, D),  $ab=ff$ for the fast-fast correlation (B, E), and $ab=fs$ for the fast-slow correlation (C, F).
}
\end{figure}

Further analyses with pair-correlation functions are informative. The pair-correlation functions of position, i.e., the radial distribution functions of fast and slow nucleosomes are
\begin{eqnarray}
g^{ab}(r)=\left[\xi_{rr}^{ab} (r)\right]_{\rm cell}/(2\pi rC^ab ), 
\end{eqnarray}
with
\begin{eqnarray}
\xi_{rr}^{ab}(r)=\left< \delta(r-|{\bf r}_i (t)-{\bf r}_j (t)|)\right>_{i\in a,j\in b},
\end{eqnarray}
where $a$ and $b$ are labels for either fast or slow nucleosomes, and $C^{ab}$ is a constant to normalize the function as $g^{ab}(r)\rightarrow 1$ for large enough $r$. In Eq.~4, $\left<\cdots \right>_{i\in a,j\in b}$ is the average over $t$ and nucleosomes of type $a$ and $b$. The nucleosome pair $i$ and $j$ were chosen to be averaged over in Eq.~4 only when both of them were observed in the same imaging frame.

Due to the small number of sampled nucleosome pairs, a pair of nucleosomes were only infrequently observed in their mutual vicinity; therefore, as shown in Figs.~4A--4C, $g^{ab}(r)$ is small for $r\lesssim 200$~nm. However, $g^{ab}(r)$ has peaks at $r\approx D^{ss}$ and $2D^{ss}$ with $D^{ss}=600$~nm (Fig.~4A), $r\approx D^{ff}$ and $2D^{ff}$ with $D^{ff}=380$~nm (Fig. 4B), and $r\approx D^{fs}=(D^{ff}+D^{ss})/2$ and $r\approx D^{fs}+D^{ss}$ (Fig.~4C); a straightforward interpretation  is that fast and slow nucleosomes constitute domains with diameter $D^{ff}$ and $D^{ss}$, respectively, and the oscillatory pattern of $g^{ab}(r)$ reflects liquid-like spatial arrangements of these domains.
This domain picture is consistent with the pair-correlations of displacement direction, ${\bf \hat{v}}_i={\bf v}_i/|{\bf v}_i|$, calculated as
\begin{eqnarray}
\xi^{ab}(r)=\left[ \xi_{vv}^{ab}(r) \right]_{\rm cell} /\left[ \xi_{rr}^{ab}(r) \right]_{\rm cell},
\end{eqnarray}
with
\begin{eqnarray}
\xi_{vv}^{ab}(r) = < {\bf \hat{v}}_i(t)\cdot{\bf \hat{v}}_j(t)\delta(r-|{\bf r}_i(t)-{\bf r}_j(t)|) >_{i\in a, j\in b}.
\end{eqnarray}
Though $\xi^{ab}(r)$ is an oscillating function of $r$, its oscillation profile is represented by $|\xi^{ab}(r)|$ as shown in Fig.~4D--4F. Correlation shown in  $|\xi^{ab}(r)|$ is large only within a certain range $r<{R_c}^{ab}$. When we define ${R_c}^{ab}$ as $|\xi^{ab}({R_c}^{ab})|=0.2$, we find that ${R_c}^{ss}\approx 300$~nm (Fig. 4D) and ${R_c}^{ff}\approx R_c^{fs}\approx 190$~nm  (Fig. 4E and 4F), showing ${R_c}^{aa}=D^{aa}/2$ and ${R_c}^{fs}=\min({R_c}^{ss},{R_c}^{ff})$, which implies that domains with radii ${R_c}^{ff}$ and ${R_c}^{ss}$ are domains of fast and slow nucleosomes, respectively, within which nucleosome dynamics are correlated with each other. We refer to these domains as fast dynamic domains (f-domains) and slow dynamic domains (s-domains). 
Comparing ${R_c}^{ff}$ and ${R_c}^{ss}$ with the radius distribution observed in the FISH measurements \cite{Boettiger2016}, the size of  f-domains is estimated as $\sim 50$--300~kb and that of  s-domains is $\sim 150$--500~kb, suggesting that the  size of f-domains is around the median size 185~kb of loop domains \cite{Rao2014} and the size  of s-domain is near to that of clusters of loop domains or TADs \cite{Dekker2015}. A clear oscillatory behavior of $g^{fs}(r)$ in Fig.~4C shows that f-domains and s-domains form a mosaic arrangement. The peak of $g^{fs}(r)$ at $r\approx D^{fs}$ shows density correlation between the adjacent f- and s-domains and the peak at $r\approx D^{fs}+D^{ss}$ implies correlation between f-domains and the next nearest s-domains. Dominance of $g^{fs}(r)$ at $r\approx D^{fs}+D^{ss}$ over $D^{fs}+D^{ff }$ suggests that f-domains are minor components in the mosaic arrangement.

We note that the displacement correlation function $\left[\xi_{vv}^{ab}(r)/\xi_{rr}^{ab}(r)\right]_{\rm cell}$  shows micrometer-scale correlations (\textcolor{blue}{Fig.~S6}) in consistency with the microscopically observed long-length scale correlations \cite{Zidovska2013, Shaban2019, Shaban2018}. However, such long-range correlations disappear when the numerator $\xi_{vv}^{ab}(r)$ is averaged over multiple cells as in Eq.~5. The correlation remaining after this averaging over different cells is the core correlation common to those cells; as shown in Figs.~4D--4F, this correlation is large only within the range $r<R_c^{ab}$. Thus, the present analyses of correlation functions provide a consistent picture that fast and slow nucleosomes show fluid movement to form dynamically correlated regions, i.e., f- and s-domains.

\subsection*{Perturbations on chromatin movement}
The relationship between chromatin dynamics and the physical features of domains is further examined by comparing the movement of nucleosomes under different cell conditions. Nozaki et al.~\cite{Nozaki2017} observed single-nucleosome movement in the following cases. (i) {\bf Cohesin knockdown (KD)}: cohesin action was suppressed by siRNA knockdown of a cohesin subunit RAD21 \cite{Nasmyth2005, Shintomi2010}, which diminished  the frequency of  chromatin chain to be bundled by cohesin. (ii) {\bf Histone tail hyper-acetylation}: the frequency of histone tail acetylation was globally increased by adding a histone deacetylase inhibitor, Trichostatin A (TSA). Histone tail acetylation leads to weakening the histone H3 and H4 tail binding to the neighboring nucleosome and subsequent decondensation of chromatin \cite{Gorisch2005}. (iii) {\bf Crosslinking of chromatin}: chromatin chains were crosslinked by treating the cells with formaldehyde (FA). Also compared was the case of (iv) {\bf focusing on heterochromatin:} changing the height of microscopic focal layer from the center to the periphery (PERI) of the nucleus, where these regions are enriched with heterochromatin regions or lamina-associated domains (LADs) \cite{Steensel2017}, tethered to inner nuclear membrane proteins \cite{Lemaitre2015}.

\begin{figure}
\centering
\includegraphics[width=0.5\linewidth]{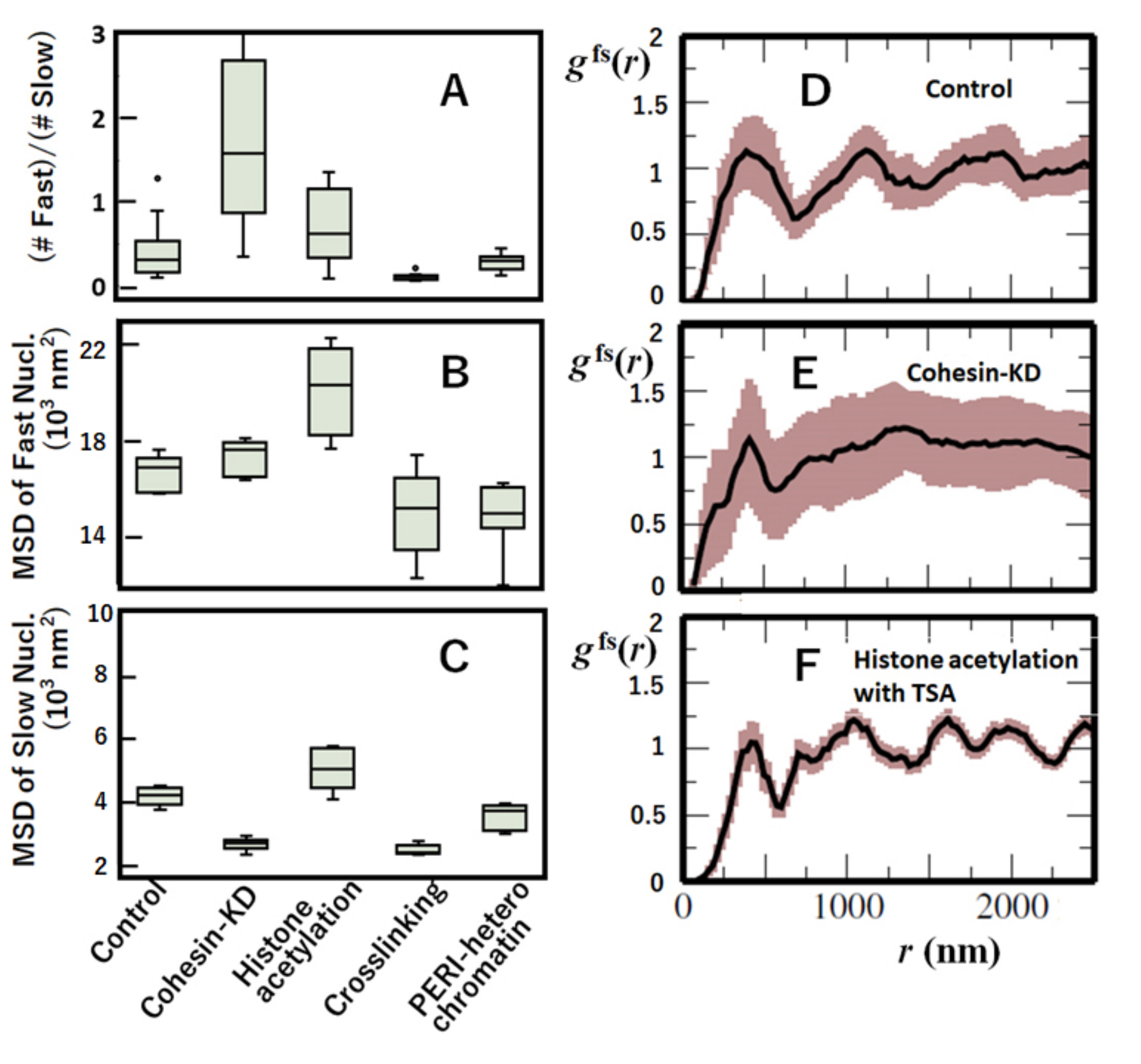}
\caption{
Effects of perturbations on cells and effects of focusing on heterochromatin. (A--C) Features of the effects on the distribution of mean square displacement (MSD), $P(M,t)$ at $t=0.5$~s, of single nucleosomes; (A) the ratio of the number of fast nucleosomes to the number of slow nucleosomes, (B) the mean MSD of fast nucleosomes, and (C) the mean MSD of slow nucleosomes. Box plots of the data from 10 cells. (D--F) The radial distribution function $g^{fs}(r)$ in cases of (D) control, (E) cohesin-KD, and (F) histone hyper-acetylation with TSA. In D--F, width of brown shaded area shows the standard errors among 10 cells.
}
\end{figure}
The distribution $P(M,t)$ in these cases is bimodal or multimodal (\textcolor{blue}{Fig.~S7}), so that the fast and slow nucleosomes are defined in the same way as in control described in the previous subsections. Features of $P(M,0.5~{\rm s})$ are summarized in Figs.~5A--5C. The ratio of the number of fast nucleosomes to the number of slow nucleosomes is markedly large in cohesin-KD and small in FA-crosslinking cases. The average MSD for the fast nucleosomes is large in histone tail acetylation with TSA, and the average MSD for the slow nucleosomes is small in cohesin-KD and FA.

These features can be understood when chromatin domains are modified in a particular way by each perturbation. Cohesin-KD diminishes the cohesin action to bundle the chromatin chains \cite{Nasmyth2005, Shintomi2010}, decreasing the constraint on the chain movement. Loosening the motional constraint increases the population of fast nucleosomes. It is intriguing to see that with cohesin-KD the slow nucleosomes become slower and fast nucleosomes become faster, which may be due to the enhancement of  A/B compartmentalization as found in the enhanced contrast of the Hi-C contact pattern upon cohesin depletion \cite{Schwarzer2017, Rao2017}. 
Adding TSA, on the other hand, causes global decondensation of the compact domains; this structural loosening makes fast and slow nucleosomes faster. By contrast, FA-crosslinking induces constraints on the movement, which severely decreases the population of fast nucleosomes and also decreases the average MSD of fast and slow nucleosomes. The milder but similar effect is found in PERI-heterochromatin, which should reflect the nucleosome tethering to the nuclear lamina and other heterochromatin proteins \cite{Steensel2017,Lemaitre2015}.

With TSA, the first peak of the radial distribution function $g^{ff}(r)$ increases (\textcolor{blue}{Fig.~S8}), which is consistent with the enhancement of fast nucleosomes with TSA. Shown in Figs. 5D--5F is the  function $g^{fs}(r)$.
With TSA, $g^{fs}(r)$ has a shorter length scale of oscillation than in control. By contrast, the oscillation in $g^{fs}(r)$ diminishes with cohesin-KD. Thus, different dynamic constraints give different effects on the f- and s-domain  arrangement. A possible mechanism is that with TSA, s-domains are dissolved and f- and s-domains are mixed, shortening the length scale of the $g^{fs}(r)$ oscillation, while  with cohesin-KD, the A/B compartmentalization is enhanced, which separates f- and s-domains,
diminishing the mosaic-like arrangement and suppressing the $g^{fs}(r)$ oscillation. To examine this possibility, further comparison between changes in dynamic features and  Hi-C contact maps upon cell perturbations is desired.

The above analyses showed that the constraints on the motion of domains slow down the movement; the cohesin bundling of chromatin chains, which is diminished by cohesin-KD, and the intra-domain nucleosome-nucleosome interactions that decrease on the addition of TSA, are the effective constraints on the movement. 
This supports a view that nucleosomes are driven primarily by thermal fluctuating motion and physical or geometrical constraints on the motion are responsible for separating slow nucleosomes from the fast ones.

\section*{Discussion}
The present statistical analyses showed that physical or geometrical constraints on the motion  are responsible for separating slow nucleosomes from the fast ones. Indeed, as shown in the TSA treated cells, decondensing chromatin increased the mobility of nucleosomes, and as shown in PERI-heterochromatin, tethering of chromatin to lamina and other proteins slows the domain movement. In addition to these factors described above, transcription machinery is likely to be another important factor. As well as the classical transcription factory model \cite{Feuerborn2015}, the recent observations of the droplet-like assembly of RNA polymerase II and transcription coactivators/factors \cite{Lu2018, Cho2018, Boija2018, Sabari2018} suggested that enhancers/promoters can bind to such clusters/droplets  to form a loosely connected network of chromatin chains, which slows the movement of transcriptionally active chromatin regions \cite{Nagashima2019}. Importantly, slowing down of the nucleosome movement is not induced by a sole mechanism but there are multiple coexisting mechanisms; therefore, the interplay between these mechanisms
 is essential to understand the origin of the slow movement.

Here, we introduce a minimal model to discuss the interplay of multiple dynamical constraints. As illustrated in Fig. 6A, two consecutive loop domains with cohesin bound at their boundaries are represented by a ring having two regions, Region I and Region II. This ring is a bead-and-spring chain consisting of 300 beads with each bead representing a 
$\sim 1$~kb segment. Region I consists of 100 beads and Region II consists of the rest 200 beads, and two regions are separated by a contact between the 1st and 101st beads, which mimics the cohesin binding. We simulate movement of this chain by numerically integrating the Langevin dynamics by assuming the interaction potential, $V(r_{ij})=\epsilon_0(r_0/r_{ij})^{12}-\epsilon_{ij} (r_0/r_{ij})^6$, where $r_{ij}$ is distance between $i$ and $j$th beads, $r_0$ is a unit length, and we set $\epsilon_0/(k_{\rm B}T)=1$. See the \textcolor{blue}{SI Text} for more details. The coefficient for the attractive part of the potential is $\epsilon_{ij}=\epsilon_{\rm I}$ when $i\, {\rm and}\,j\in {\rm Region\,I}$, $\epsilon_{ij}=\epsilon_{\rm II}$ when $i\, {\rm and}\,j\in {\rm Region\,II}$, and $\epsilon_{ij}=0$ when $i$ and $j$ belong to different regions. In \textcolor{blue}{Fig.~S9}, We show the dependence of the simulated radius of gyration of the ring on the interaction strength of $\epsilon_{\rm I}=\epsilon_{\rm II}=\epsilon$. This ring is open extended when $\epsilon/k_{\rm B}T < 0.6$  and compact condensed when $\epsilon/k_{\rm B}T >1.2$, showing a continuous coil-globule transition as in the microscopic observation of chromatin \cite{Zinchenko2018}. We examine $\epsilon_{\rm I}$ and $\epsilon_{\rm II}$ around this transition regime. We represent the effects of tethering or interactions with the other nuclear structures by defining a reference point on the chain as the locus to be tethered. Though tethering/untethering can dynamically switch in cells, we here use a simple assumption that the coordinate system is fixed with its origin on a specific bead (reference point) and we monitor the MSD of the simulated polymer by using this coordinate system.

\begin{figure}
\centering
\includegraphics[width=0.6\linewidth]{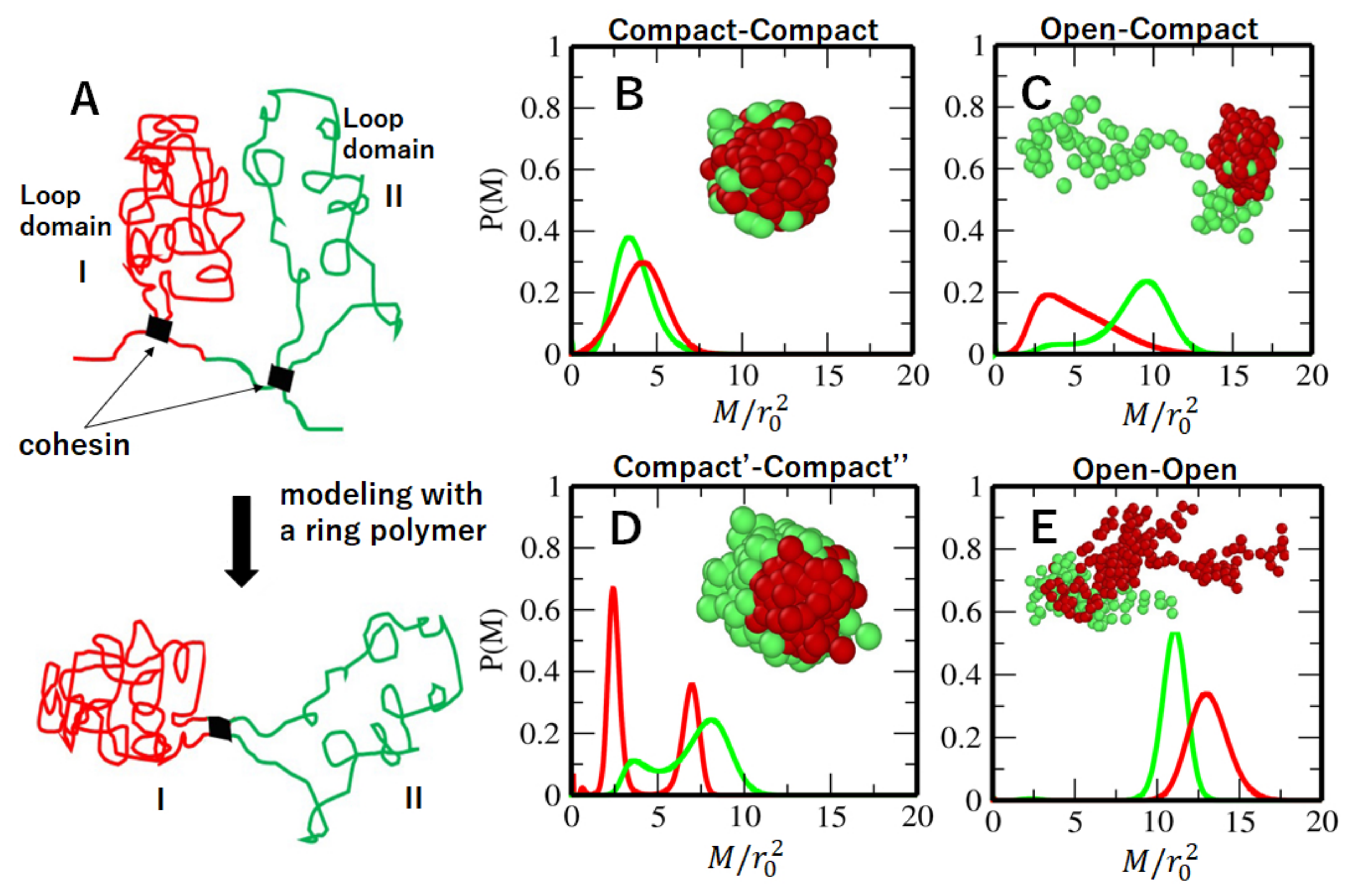}
\caption{
A polymer model of  looped domains. (A) Two consecutive looped domains 
are represented by Region I and Region II in a model ring. The cohesin binding is represented by thick bars. (B--E) Distribution of mean square displacement, $P(M)$, of beads in a polymer model. Connected two looped domains of 
(B) compact~($\epsilon_{\rm I}/k_{\rm B}T=1.0$)--compact~($\epsilon_{\rm I}/k_{\rm B}T=1.0$), 
(C) compact~($\epsilon_{\rm I}/k_{\rm B}T=1.0$)--open~($\epsilon_{\rm II}/k_{\rm B}T=0.6$), 
(D) compact'~($\epsilon_{\rm I}/k_{\rm B}T=1.2$)--compact''~($\epsilon_{\rm II}/k_{\rm B}T=0.9$), and 
(E) open~($\epsilon_{\rm I}/k_{\rm B}T=0.6$)--open~($\epsilon_{\rm II}/k_{\rm B}T=0.6$) regions. 
In B--E, $P(M)$ calculated from the reference point in Region I (red) and from the reference point in Region II (green) are plotted. Insets are snapshots of the polymer ring; beads in Region I (red) and those in Region II (green) are shown with spheres. 
}
\end{figure}

Figs. 6B--6E show $P(M)=P(M,t_m )$ obtained at a given time $t_m$ (\textcolor{blue}{SI Text}) from the simulated movement of the chain. Two regions having interactions with $\epsilon_{\rm I}/k_{\rm B}T=\epsilon_{\rm II}/k_{\rm B}T=1.0$ represent two compact loop domains (Fig.~6B). These two regions tend to have a merged condensed configuration, and $P(M)$ has a peak at a small $M$. Two regions with $\epsilon_{\rm I}/k_{\rm B}T=1.0$ and $\epsilon_{\rm II}/k_{\rm B}T=0.6$ represent the connected compact and open loop domains (Fig.~6C). When the reference point is in the compact domain of Region I, both two regions show the slow movement, while when the reference point is in the open domain of Region II, both two regions show the fast movement. Therefore, the movement depends not only on whether the loop domain is open or compact but also on the nature of the domain tethering. In a ring with $\epsilon_{\rm I}/k_{\rm B}T=1.2$ and $\epsilon_{\rm II}/k_{\rm B}T=0.9$, both two loop domains take compact configurations (Fig.~6D). However, Region I is a core globule and Region II wraps the surface of Region I. In this case, $P(M)$ shows a bimodal peak with the slow movement of Region I and the fast movement of Region II. 
The sensitivity of $P(M)$ to tethering indicates that nucleosomes belonging to compact regions can be inferred as fast. When two loop domains are open with $\epsilon_{\rm I}/k_{\rm B}T=\epsilon_{\rm II}/k_{\rm B}T=0.6$, both of them show the fast movement regardless of the position of the reference point. 
When two compact loop domains merge as shown in Fig 6B, their motions are correlated to form a single s-domain having an effectively larger size than independent loop domains, making $R_c^{ss}$ large. 
It is intriguing to  examine with the present polymer model whether the transitions between f- and s-domains take place as   open-closed structural transitions  suggested by a chromosome  model  \cite{Zhang2015}.

Thus, the minimal polymer model elucidates the interplay of key mechanisms such as geometry (compact or open, and core or surface) of the chromatin chain and tethering in their role in the f- and s-domain organization. Further quantitative analyses are necessary to identify the precise molecular interactions which define chromosome organization geometry and tethering in living cells. The statistical analyses of single-nucleosome trajectories with the enhanced sampling \cite{Nagashima2019} will allow for such quantitative analyses. As noted by Dubochet \cite{Dubochet2001}, chemical fixation such as with formaldehyde or glutaraldehyde, which is a standardized methodology in cell biology analyses, can have artifactual effects on chromatin interactions. To explore the ``in vivo'' organization, investigating chromatin in living cells is crucial.  As demonstrated in the present study, the statistical analyses of single-nucleosome trajectories provide a means for such exploration.

\section*{Conclusions}
We investigated heterogeneity of chromatin dynamics in living human cells by analyzing single-nucleosome movement. The obtained nucleosome MSD distribution revealed that the nucleosomes are categorized into two types, fast and slow.  
This categorization of nucleosomes revealed aspects of organization of heterogeneous chromatin domains; nucleosome movements are correlated with each other within f- and s-domains. This analysis was applied to cells under various perturbations, and together with a simple polymer model, the method gave a consistent picture of organization of dynamic chromatin domains. Thus, the categorization of fast and slow nucleosome movement introduced in the present analyses provides a basis for understanding chromatin organization.

\section*{Methods}
$P(M,t)$ of Eq.\,1 was calculated in an iterative way with the Richardson-Lucy (RL) algorithm:\,starting from the initial distribution, 
\begin{eqnarray}
P^1(M,t)=1/M_0\exp(-M/M_0), \nonumber  
\end{eqnarray}
$P^{n+1}(M,t)$ at the $n+1$th iteration was obtained by
\begin{eqnarray}
P^{n+1}(M,t)=P^n(M,t)\int \left[G_s(r,t)/G_s^n(r,t)\right]q(r,M)d^2{\bf {\rm r}}, \nonumber
\end{eqnarray}
with $G_s^n(r,t)=\int P^n(M,t)q(M,t)dM$. 
This equation was iterated under the constraints, $P^{n+1}(M,t)\ge 0$ and $\int P^{n+1}(M,t)dM =1$.

\section*{Acknowledgment}
This work was supported by JST CREST Grant Number JPMJCR15G2, the Riken Pioneering Project, JSPS KAKENHI (Grant Number JP19H01860, 19H05258, JP16H04746, 16H06279 (PAGS), 19H05273), Takeda Science Foundation, and NIG JOINT (2016- A2 (6)).

\bibliography{ssashwin}

\section*{Author contributions statement}
Author contributions: SSA, TN, KM and MS designed  research and wrote the manuscript. SSA developed  methods and performed research, and SSA and MS analyzed the data.

\section*{Additional information}
The authors declare no conflict of interest.

\end{document}